# A Hybrid Quantum-Classical Approach for Melt Pool Prediction in Laser Powder Bed Fusion

**Matthew M. Sato and Kincho H. Law**
Engineering Informatics Group
Department of Civil and Environmental Engineering
Stanford University
Stanford CA, 94305

## Abstract

*Laser powder bed fusion (LPBF) is a promising additive manufacturing technique that suffers from quality assurance concerns. Predicting melt pools from process parameters is crucial for assessing quality prior to manufacturing but remains a difficult problem because of the complex physical processes underlying LPBF. Quantum computers present a new computing paradigm, providing a new approach to information processing using quantum entanglement and superposition. This paper presents a practical demonstration of a hybrid quantum-classical model that leverages quantum computing to improve process parameter feature extraction with a quantum feature encoder. To make the quantum approach computationally feasible for large datasets, we first employ a clustering algorithm to reduce the number of expensive quantum computations. These quantum features are then processed by a classical neural network to predict the melt pool morphology, allowing for more accurate predictions of melt pools. We demonstrate the method using a quantum simulator, analyze the effect of measurement shot noise on the predictive performance of the network, and verify the results using quantum hardware. Finally, by examining which quantum features are most important, we provide insights that can inform the future design of more effective quantum encoding circuits. Ultimately, the performance improvement over purely classical networks validates the hybrid approach, demonstrating an engineering application of quantum computing using noisy and intermediate scale quantum (NISQ) devices.*

## 1. Introduction

Laser powder bed fusion (LPBF) is an important additive manufacturing (AM) technique that enables the fabrication of unique, geometrically complex parts that are often impossible or uneconomical to produce with traditional manufacturing techniques. The success of the LPBF process, however, is dependent on a complex relationship between the process parameters (machine settings such as laser power and scan speed) and the final part quality (often measured by strength and geometrical accuracy) [1]. The LPBF process (Figure 1) uses a high-power laser to selectively sinter metal powder into the desired shape on a layer-by-layer basis. The extreme temperature gradients generated by the laser and rapid material state changes mean that poorly chosen process parameters can lead to detrimental defects, such as lack of fusion and keyhole porosity [2], compromising the part's mechanical strength.

A promising approach for quality assurance is through in-situ monitoring of the melt pool [3,4], the region of molten metal around the laser. The melt pool morphology as captured by a co-axial camera can vary significantly, as shown in Figure 2, and serves as a real-time indicator of the LPBF quality based on shape, size, or intensity. Predicting the melt pool shape from a set of process parameters is a difficult, high-dimensional mapping problem. While high-fidelity physics-based simulations can model the LPBF process, they are computationally prohibitive for real-time or large-scale analysis.

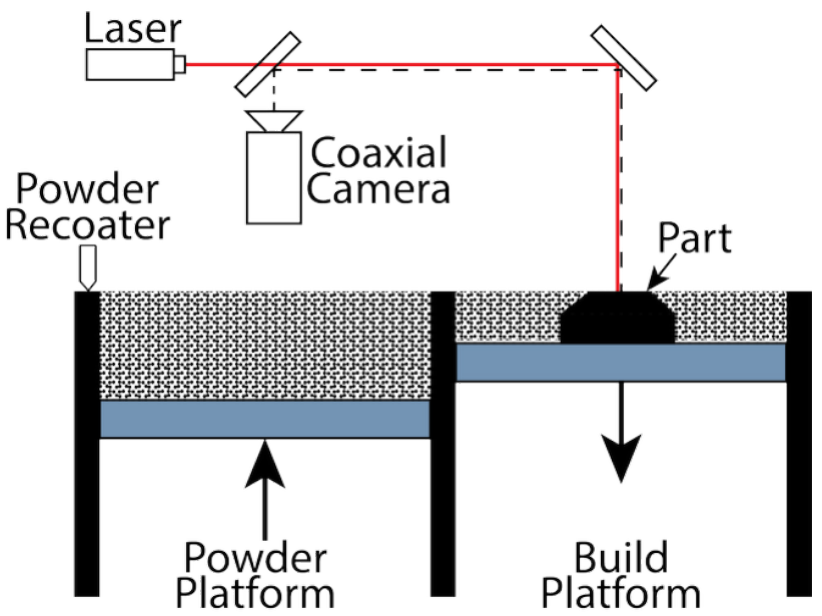

**Figure 1:** A simplified diagram of the LPBF process with co-axial camera measurement.

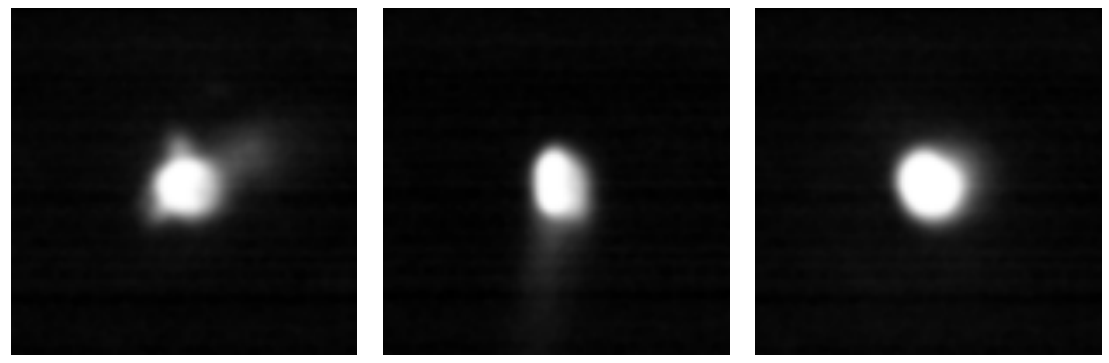

**Figure 2:** Examples of melt pool images created by different process parameters and captured by a co-axial camera.

The computational difficulty of LPBF modeling has motivated the use of data-driven machine learning models, but the inherent complexity of the process still poses a challenge for purely classical machine learning models. To address this challenge, we use the emerging paradigm of quantum computing [5,6] to explore data-driven techniques that are better able to predict LPBF melt pool characteristics from process parameters.

Unlike classical computers, which process information in bits, quantum computers leverage the principles of superposition and entanglement to manipulate qubits in the high-dimensional Hilbert space [7]. Quantum computers provide a fundamentally new way to represent and process complex data. While classical machine learning techniques have shown great promise in the LPBF domain [8], we hypothesize that the unique and expressive powers of the quantum information feature space can create more effective representations of LPBF data, improving the performance of melt pool prediction models.

This work is grounded in the limits of current quantum hardware, which exists in the Noisy Intermediate Scale Quantum (NISQ) era. Current state-of-the-art quantum computers are characterized by a limited number of qubits ($\approx 100$) and are susceptible to decoherence and gate errors from environmental noise and imperfect quantum control. As such, NISQ devices cannot yet execute fault-tolerant algorithms that provide exponential speedups (such as Shor's factoring algorithm) but can still provide advantages over classical computers [9]. Therefore, instead of pursuing a quantum speedup, our investigation focuses on harnessing the unique representational power of NISQ devices. We propose employing a NISQ processor as a specialized co-processor to generate a rich feature representation for a classical neural network, demonstrating a tangible application for current quantum hardware in a practical engineering domain.

In this paper, we propose and demonstrate a hybrid quantum-classical neural network for predicting the LPBF melt pool from process parameters. Our methodology uses a quantum circuit to encode the classical process parameters into a quantum state, from which an improved feature vector can be measured. This quantum-derived feature vector is then fed into a classical neural network. We show that by first mapping the process parameters into a high-dimensional quantum feature space, the predictive performance of the classical neural network is improved. Furthermore, we examine the encodings created by the quantum encoder and investigate the effect of finite-shot quantum measurements, allowing us to explore the trade-off between the quantum computational cost (number of shots) and prediction accuracy. The primary goal of this work is to demonstrate and explore an early use case for NISQ-era quantum computers as applied to data-driven methods in AM.

In summary, the contributions of this paper are:

- an exploratory procedure for encoding LPBF process parameters using a quantum computer,
- a hybrid quantum-classical neural network architecture tailored for LPBF melt pool prediction,
- a demonstration of the proposed method using both simulated and real quantum hardware on an experimental melt pool dataset,
- and an examination of the effect of measurement shots on prediction performance.

The remainder of this paper is organized as follows: Section 2 reviews related work. Section 3 provides the necessary background on quantum computing and details our methodology. Section 4 presents the experimental results on a melt pool dataset. Finally, Section 5 provides concluding remarks and potential future work.

## 2. Related Works

This work sits at the intersection between AM process control and quantum machine learning. We first review the state-of-the-art in LPBF process monitoring and modeling, identifying both physics-based and classical machine learning techniques. Then, we provide relevant work using quantum computing in machine learning, highlighting techniques and existing applications of quantum computers.

### 2.1 LPBF Process Monitoring and Defect Detection

The pursuit of part quality and reliability in LPBF has driven significant research into in-situ monitoring. The formation of defects in LPBF are governed by complex physical phenomena within the melt pool and are influenced significantly by the process parameters [10]. Two of the most common defects are lack of fusion porosity and keyhole porosity [2]. Lack of fusion porosity occurs from insufficient laser energy density, leading to incomplete melting between scan tracks or layers. In contrast, keyhole porosity results from a deep vapor depression, which causes the melt pool to collapse and trap gas within the final part. Researchers have proposed a wide variety of sensing devices aimed at monitoring the process, including ultrasonic, acoustic, optical, x-ray, pyrometry, and infrared sensors [3].

### 2.2 Classical Approaches for LPBF Process Modeling

To interpret data and optimize process parameters, two primary computational approaches are used: physics-based simulation methods and data-driven machine learning methods. While high fidelity physics-based methods provide insights into melt pool fluid dynamics [11], they are computationally prohibitive for real-time applications or restricted to small parts. Machine learning provides a more computationally tractable alternative driven by the data collected from process monitoring. Researchers have effectively used machine learning methods to classify/predict melt pools and to optimize process parameters [12–14]. While machine learning is effective, we investigate how quantum computers can improve upon these purely classical approaches.

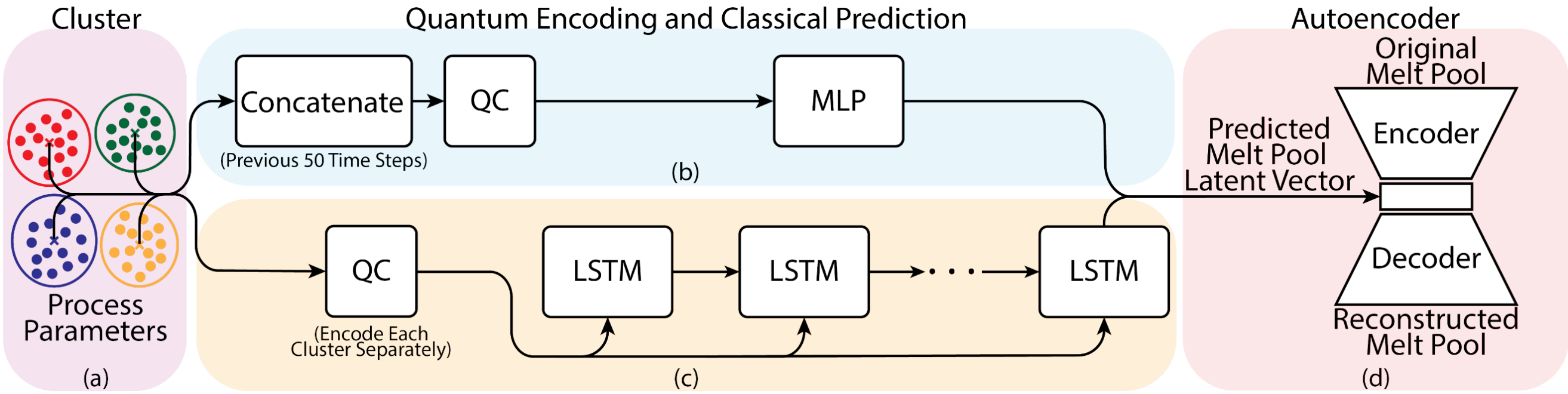

**Figure 3:** Overview of the melt pool prediction: (a) clustering to group process parameters, (b) a quantum circuit (QC) and fully connected predictor, (c) a QC and long short-term memory predictor, and (d) a convolutional autoencoder for melt pool latent representation.

### 2.3 Quantum Computing and Machine Learning

Quantum machine learning techniques seek to leverage quantum computers for data analysis tasks [15]. Quantum machine learning approaches use quantum feature maps [16] or variational quantum algorithms (VQAs) [17], either separately or in combination. Quantum feature maps are used to encode classical data into the quantum feature space using a fixed quantum circuit design and are stable due to their fixed parameters, but their capabilities are reliant on the quantum circuit selection. Meanwhile, VQAs use a classical optimizer to learn parameters of a quantum circuit. VQAs have shown promise for a variety of applications [18,19] but are susceptible to the barren plateau problem [20] where the massive quantum state space leads to vanishing gradients which make optimization difficult. Various applications have been demonstrated using quantum computing, for example using a quantum feature map for estimating fill probabilities in finance [21] and using a quantum convolutional neural network (a VQA) for detecting spatter in LPBF [22]. Inspired by quantum feature maps in finance applications [21], this work adopts the quantum feature map approach for encoding LPBF process parameters, leveraging the representational power of quantum states while avoiding the training challenges associated with VQAs.

## 3. Methods

This section outlines our hybrid quantum-classical machine learning approach for predicting melt pools from LPBF process parameters. The methodology consists of the stages shown in Figure 3. First, we use a clustering strategy to reduce the number of quantum circuit executions. Second, a quantum circuit maps the LPBF process parameters into a feature vector. Third, a classical neural network uses the quantum encoded feature vector to predict a low-dimensional latent representation of the melt pool image. Last, a convolutional autoencoder decodes the predicted latent vector to produce the final predicted melt pool. Before detailing the specific methodology, we provide the necessary requisite background on quantum computing.

### 3.1 A Primer on Quantum Information

To understand the quantum techniques used in this paper, we provide a brief introduction to quantum information. For a comprehensive review, refer to Nielsen and Chuang [7].

#### *3.1.1 Qubits and Superposition*

The fundamental unit of quantum information is the qubit. Unlike a classical bit, which is always in a discrete state of $0$ or $1$, a qubit exists in a linear combination of states, called superposition:

$$|\psi\rangle = \alpha|0\rangle + \beta|1\rangle \tag{1}$$

where $|\psi\rangle$ is a quantum state and $\alpha, \beta \in \mathbb{C}$ are the complex probability amplitudes of the $|0\rangle$ and $|1\rangle$ states satisfying the normalization constraint: $|\alpha|^2 + |\beta|^2 = 1$. A qubit's state can be visualized as a point on the surface of the Bloch sphere (Figure 4). Although this paper uses the standard (computational) basis states ($|0\rangle/|1\rangle$), a qubit can be expressed in any orthonormal basis (i.e., $|+\rangle/|-\rangle$ and $|+i\rangle/|-i\rangle$).

#### *3.1.2 Multi-Qubit Systems and Entanglement*

The computational advantage of quantum systems lies in the exponential scaling of the state space. An $n$-qubit system exists in the $2^n$-dimensional Hilbert space, requiring $2^n$ complex amplitudes to describe the state. The exponential scaling results in many parameters for even a small number of qubits, allowing for the representation of complex information. An important feature of multi-qubit systems is entanglement, where qubits become correlated in a way that is impossible classically. Multi-qubit systems are considered entangled if the system *cannot* be written as a tensor product of individual qubit states ($|\psi\rangle \neq |\psi_1\rangle \otimes |\psi_2\rangle$). For example, the Bell states,

$$|\Phi^{\pm}\rangle = \frac{1}{\sqrt{2}}(|00\rangle \pm |11\rangle) \tag{2}$$

are examples of entangled qubits. Entangled qubits behave as a system, regardless of their physical separation, and measuring one qubit instantly influences the state of the entangled qubits.

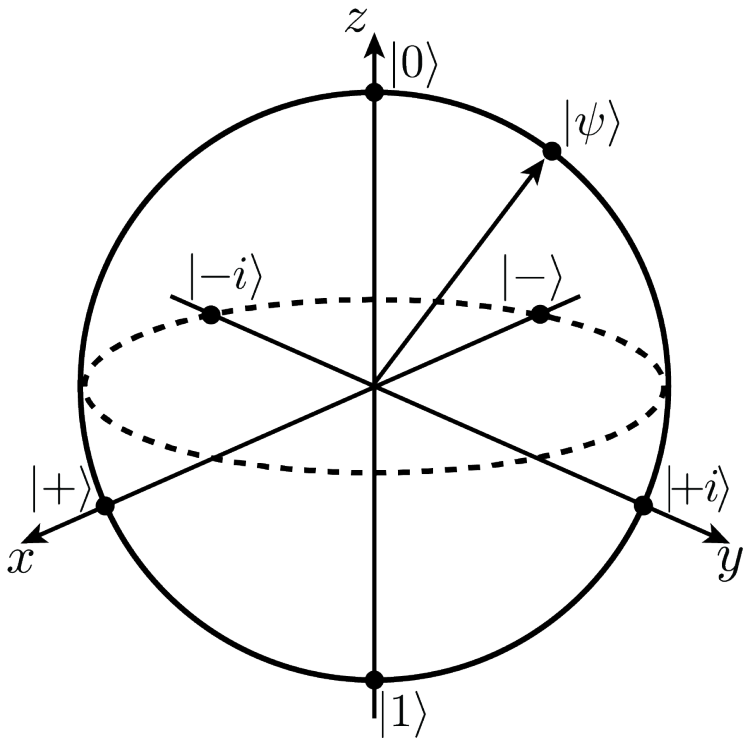

**Figure 4:** The Bloch sphere for a single qubit.

*3.1.3 Quantum Gates and Quantum Circuits*

Quantum computations are performed by applying quantum gates, which are represented as unitary operators, $U$. A unitary operator preserves the normalization of the state vector and is a rotation on the Bloch sphere for a single qubit. In this paper, we use parameterized single-qubit rotation gates ($R_x(\theta)$, $R_y(\theta)$, $R_z(\theta)$), and two-qubit rotation gates,

$$R_{ii}(\theta) = R_i(\theta) \otimes R_i(\theta) \tag{3}$$

where $i \in \{x, y, z\}$ and $\theta$ is the rotation angle. Quantum gates are combined in sequence to form quantum circuits, an analog of a classical logic circuit.

*3.1.4 Quantum Measurement*

Quantum information is retrieved from a qubit through irreversible measurement, where a qubit collapses to one of its basis states upon measurement. Measurements can theoretically be performed in any basis but are usually performed in the computational ($z$) basis due to physical constraints. In the computational basis, the state $|\psi\rangle$ from Eq. 1 collapses to $|0\rangle$ with probability $|\alpha|^2$ and collapses to $|1\rangle$ with probability $|\beta|^2$. To perform measurements in other bases (i.e., along the $x$ or $y$ axis of the Bloch sphere), a unitary must be applied to map the desired measurement basis to the computational basis. Because measurements are destructive and probabilistic, a single measurement does not reveal the underlying amplitudes. Instead, the true state of a qubit can only be recovered by repeatedly executing a quantum circuit to compute expectation values for the underlying amplitudes.

### 3.2 Quantum Feature Encoder

To improve neural network performance, we encode the process parameters using a quantum circuit that acts as a quantum feature map. The quantum circuit is inspired by the time evolution of an $n$-qubit quantum system according to the solution of the time-dependent Schrödinger equation:

$$|\psi(t)\rangle = \exp(-iHt/\hbar)|\psi(0)\rangle \tag{4}$$

where $t$ is time, $\hbar$ is the reduced Planck constant, and $H$ is the Hamiltonian which governs how the system evolves over time. In this work, the system evolves under the Heisenberg Hamiltonian [23]:

$$H = \frac{1}{4}\sum_{j=0}^{n-2} J_j^x \sigma_j^x \sigma_{j+1}^x + J_j^y \sigma_j^y \sigma_{j+1}^y + J_j^z \sigma_j^z \sigma_{j+1}^z \tag{5}$$

where $\sigma_j^i$ $(i = x, y, z)$ denotes the Pauli operator acting on the $j$th qubit and $J_j^i$ are coupling constants describing interaction strength between neighboring qubits. While the Heisenberg Hamiltonian has physical meaning in describing spin interactions, we use the Hamiltonian only as a template for constructing a parameterized quantum circuit capable of generating complex correlations.

According to the solution of the Schrödinger equation in Eq. 4 and using the Heisenberg Hamiltonian, the evolution of the quantum system is described by the unitary:

$$U = \exp\left(-i\frac{\alpha}{4}\sum_{j=0}^{n-2} J_j^x \sigma_j^x \sigma_{j+1}^x + J_j^y \sigma_j^y \sigma_{j+1}^y + J_j^z \sigma_j^z \sigma_{j+1}^z\right) \tag{6}$$

where $\alpha$ is an arbitrary term representing the time and Planck constant. Eq. 6 is approximated using Trotterization [24], which decomposes the operator into products of simple gate operations:

$$U = \left(\prod_{j=\text{odd}}^{n-2} U_j \prod_{j=\text{even}}^{n-2} U_j\right)^M \tag{7}$$

where $M \in \mathbb{N}$ is the number of repetitions of the operator and $U_j = R_{xx}\left(\frac{\alpha J_j^x}{2M}\right) R_{yy}\left(\frac{\alpha J_j^y}{2M}\right) R_{zz}\left(\frac{\alpha J_j^z}{2M}\right)$ is a series of two-qubit rotation gates applied to qubits $j$ and $j+1$. The product symbol in Eq. 7 represents the sequential application of quantum gates in order from right to left and builds a circuit with two layers. The first layer acts on all even indexed qubit pairs (e.g., (0,1), (2,3)), while the second layer acts on all odd indexed pairs (e.g., (1,2), (3,4)). We use Eq. 7 as a fixed depth circuit template rather than approximating physical phenomena, so the Trotter step count $M$ is set to $1$. The parameters of this circuit, the global scaling constant $\alpha$ and local coupling strengths $J_j^i$, are not trained and are set directly by the LPBF process parameters.

Next, we design the initial state of the system, $|\psi(0)\rangle$, by choosing a fixed uniformly random state for each qubit. Since quantum systems are typically initialized to a ground state where each of the $n$ qubits are set to $|0\rangle$, two rotations $R_y(\theta)$ and $R_z(\phi)$ are applied to each qubit to allow for any possible quantum state. Thus, the final unitary is described as

$$U = \left(\prod_{j=\text{odd}}^{n-2} U_j \prod_{j=\text{even}}^{n-2} U_j\right) \otimes_j R_z(\phi_j) R_y(\theta_j) \tag{8}$$

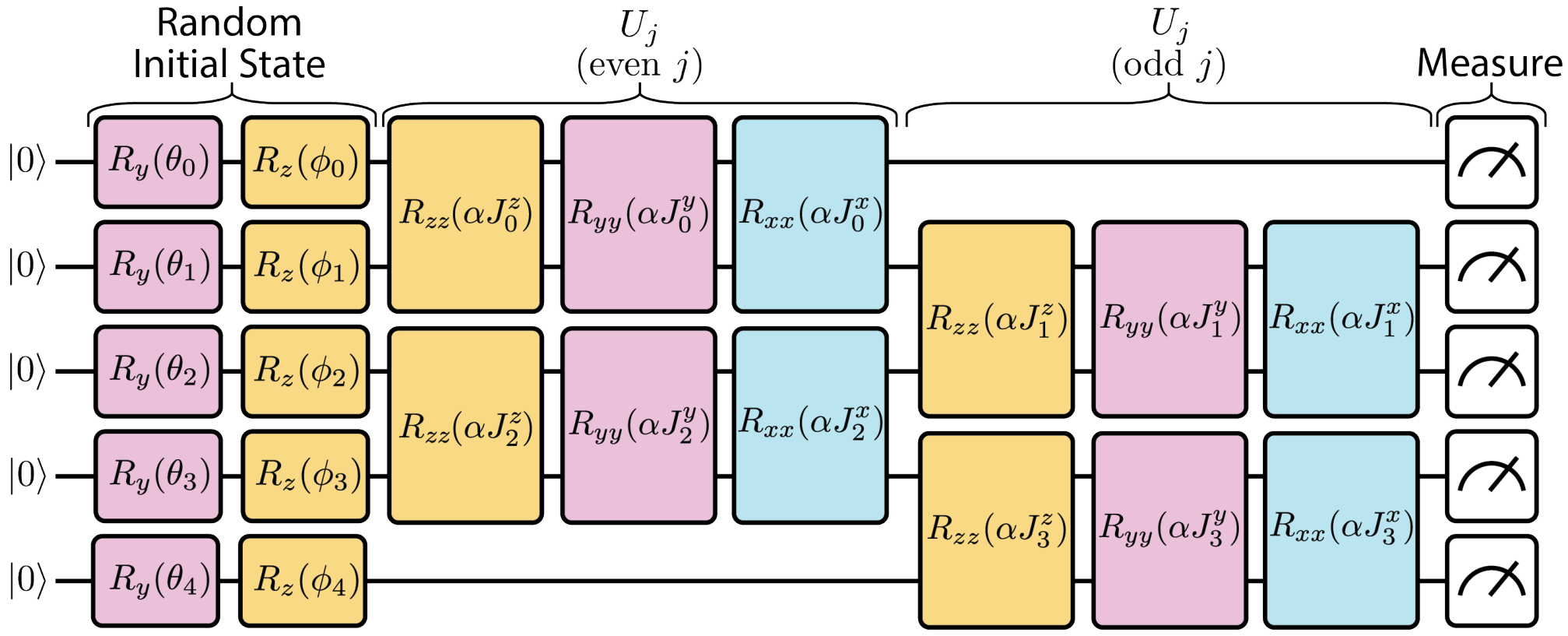

**Figure 5:** The quantum circuit for encoding melt pool process parameters using five qubits. The circuit is read chronologically from left to right.

and is shown visually in Figure 5.

After the state has evolved according to Eq. 8, the quantum state is measured to obtain the quantum feature encoding. We measure the expectation values of a set of observables $\{O_k\} = \{\sigma_j^x, \sigma_j^y, \sigma_j^z\}_{j=1}^n$, the set of single-qubit Pauli operators for each qubit. The expectation value of each observable is estimated by averaging the results over many measurement shots. The resulting quantum encoding is $x \in [-1,1]^{3n}$, a real-valued vector of length $3n$ (three observables for each qubit), where each entry corresponds to the expectation value of one Pauli observable.

### 3.3 Classical Deep Learning for Melt Pool Prediction

Following the quantum encoding of the process parameters, a classical neural network predicts the melt pool shape. To make the prediction task more tractable, we use a two-stage approach.

#### *3.3.1 Target Latent Vector Generation with an Autoencoder*

To simplify the prediction task, we avoid predicting the high-dimensional melt pool image directly. Instead, our model predicts a low-dimensional latent vector representation of the image. A classical convolutional autoencoder illustrated in Figure 3d is pretrained on the full dataset of melt pool images. The autoencoder learns to compress melt pool images into a low-dimensional latent vector and then reconstructs the image from the latent vector. The latent vector is a dense representation of the melt pool's essential morphology and serves as a target for the second stage of the melt pool predictor. The goal of our melt pool prediction model is to predict this much simpler latent vector, which can then be passed to the pretrained decoder to generate the final melt pool image.

#### *3.3.2 Deep Learning for Melt Pool Prediction*

We explore two neural network architectures for predicting melt pool latent vectors from a time-series of process parameters. Because we aim to demonstrate the feasibility of quantum computer applications for LPBF, we limit the experiments to only two simple neural networks:

- **Fully Connected Neural Network (FCNN)**: The network as shown in Figure 3b treats all time steps as a single input. All process parameters are concatenated and encoded simultaneously by the quantum circuit. The resulting single quantum feature vector is then processed by a multi-layer perceptron (MLP).
- **Long Short-Term Memory (LSTM)**: The recurrent neural network architecture shown in Figure 3c is designed to capture the temporal dependencies in the sequential process parameters. Here, the process parameters for each time step are encoded separately by the quantum circuit. The sequence of 50 quantum encoded feature vectors are then processed chronologically by the LSTM.

### 3.4 Computational Optimization via Clustering

Quantum circuit execution is expensive in both simulation and on physical quantum hardware. Circuit simulation requires significant time and memory, scaling with the number of qubits and measurement shots. Meanwhile, the scarcity of quantum processors introduces substantial monetary costs for execution on quantum hardware. To reduce the number of quantum circuit evaluations required, we apply K-means clustering to the process parameters. Rather than encoding all unique process parameter combinations, we cluster similar process parameters and encode only the cluster centers with the quantum circuit. During prediction, each process parameter combination uses its assigned cluster center's quantum encoding as input to the neural network.

We employ mini-batch K-means [25], a variant designed for large datasets, to cluster the process parameters. The number of clusters must balance two competing objectives: being large enough to capture the diversity of the process parameters while remaining small enough to achieve meaningful computational savings. This approach requires only one quantum circuit execution per cluster, a significant reduction from one quantum circuit execution per sample.

## 4. Results

To demonstrate the hybrid quantum-classical approach, the proposed methodology is demonstrated on a melt pool image dataset. The quantum encoder is implemented using both a quantum simulator and an IBM quantum computer. The results are compared to a purely classical neural network to demonstrate how a quantum computer can improve the results. Further, this section investigates the effects of different measurement shots and analyzes the quantum encodings.

### 4.1 Data

This study utilizes the publicly available "3D Scan Strategies" dataset from the National Institute of Standards and Technology (NIST) Additive Manufacturing Metrology Testbed (AMMT) [26,27]. The dataset consists of 16 experiments and provides two synchronized data streams: the time-series data of the laser's position and power recorded at 100 kHz, and images of the melt pool captured by a co-axial camera at 2 kHz. The frequency mismatch results in a sequence of 50 process parameter measurements for each captured melt pool image, forming the basis for our time-series prediction task. Examples of melt pool images taken by the coaxial camera are shown in Figure 2. In addition to laser power and position, we derive secondary parameters, including laser velocity, laser acceleration, and laser energy density (defined as laser power divided by laser velocity). The dataset is partitioned into training (80%), validation (10%), and test (10%) sets.

### 4.2 Quantum Circuit for the Different Networks

Two distinct quantum encoding strategies are used to accommodate the different input structures of the LSTM and FCNN network architectures. Both strategies map normalized process parameters to the coupling constants of the Heisenberg Hamiltonian, as described below:

- LSTM: The LSTM network processes the input data sequentially and thus each of the 50 time steps are quantum encoded individually. For a single time step, the 10 process parameters listed in Table 1 are encoded using an isotropic Hamiltonian ($J_j^x = J_j^y = J_j^z$). Each of the 10 parameters are assigned to a $J_j$ coupling constant, requiring 11 qubits. This circuit yields a 33-dimensional output vector (3 Pauli observables per qubit) for each time step.
- FCNN: The FCNN network processes the full sequence of 50 time steps simultaneously, resulting in a large input size of 350 process parameters. To manage the high-dimensional input, we use the anisotropic form of the Hamiltonian ($J_j^x \neq J_j^y \neq J_j^z$), encoding three distinct process parameters into the $J_j^x$, $J_j^y$, and $J_j^z$ constants for each adjacent qubit pair. Consequently, the quantum circuit requires 118 qubits, which is feasible on current NISQ hardware, and produces a 354-dimensional output vector.

We use a global scaling factor of $\alpha = 1$ for both experiments. Prior to encoding, each process parameter (PP) is normalized to a gate rotation angle by scaling according to the mean ($\mu$) and standard deviation ($\sigma$) of the process parameters:

**Table 1:** The process parameters and quantum encoder features.

| Process Parameter | Network Type<br>LSTM | <br>FCNN |
|---|---|---|
| Laser Power, $P$ | ✓ | ✓ |
| Change in $x$ position, $\Delta x$ | ✓ | ✓ |
| Change in $y$ position, $\Delta y$ | ✓ | ✓ |
| Laser velocity $x$ component, $v_x$ | ✓ | ✓ |
| Laser velocity $y$ component, $v_y$ | ✓ | ✓ |
| Laser velocity magnitude, $v$ | ✓ | ✓ |
| Laser energy density, $P/v$ | ✓ | ✓ |
| Laser acceleration $x$ component, $a_x$ | ✓ | |
| Laser acceleration $y$ component, $a_y$ | ✓ | |
| Laser acceleration magnitude, $a$ | ✓ | |
| **Quantum Circuit Property** | **LSTM** | **FCNN** |
| Isotropic | Yes | No |
| Qubits | 11 | 118 |
| Observables (Outputs) | 33 | 154 |
| Previous Time Steps | 50 | 50 |
| $\alpha$ | 1.0 | 1.0 |

$$\widetilde{\mathrm{PP}} = 2\pi \tanh\left(\frac{\mathrm{PP} - \mu}{3\sigma}\right) \quad (12)$$

This normalization maps the parameters into the range $(-2\pi, 2\pi)$, ensuring they are suitable as rotation angles. Each normalized parameter, $\widetilde{\mathrm{PP}}$, is assigned to one of the $J_j^{x,y,z}$ values in the quantum circuit.

### 4.3 Quantum-Classical Network Implementation

This subsection provides the implementation details for the quantum feature extractor and the classical neural networks. The source code for implementing the quantum circuit (simulation and hardware) and the neural networks is publicly available at https://github.com/satomm1/meltpool-quantum.

#### *4.3.1 Quantum Encoding Circuit*

The quantum circuit is implemented using the Qiskit software development kit (SDK) [28]. The circuit architecture is shown in Figure 5, with observables defined as the set of single-qubit Pauli operators ($\sigma_j^x$, $\sigma_j^y$, $\sigma_j^z$) for each qubit. We explore both simulated quantum results and execution on quantum hardware:

- Simulator: Ideal noiseless simulations are performed using Qiskit's Estimator primitive with the matrix product state

simulation method [28,29]. The matrix product state method is highly efficient for quantum states with limited entanglement and is well suited for our circuit, which is shallow and contains entangling gates only between neighboring qubits. To analyze the effects of statistical noise, we use the Sampler primitive [28], which mimics the process of repeated measurement shots.

- Measurement: Since quantum hardware typically measures in the computational ($z$) basis, measuring the $\sigma_j^x$ and $\sigma_j^y$ observables require basis transformations before measurement. To measure $\sigma_j^x$, a Hadamard gate is applied just prior to measurement. To measure $\sigma_j^y$, a $S^\dagger$ gate ($-90$ degree rotation about the $z$-axis) and a Hadamard gate are applied. The expectation value for each observable is then computed from the distribution of $|0\rangle$ and $|1\rangle$ measurement outcomes.
- Hardware Execution: Quantum hardware experiments are conducted on the 133-qubit IBM Torino processor, a Heron r1 device [30]. The built-in Pauli Twirling error mitigation technique is used to mitigate inherent hardware noise [31].

*4.3.2 Classical Neural Networks*

The classical component of our model consists of three neural network components: a convolutional autoencoder (AE), long short-term memory RNN (LSTM), and fully connected neural network (FCNN). The AE is used for dimensionality reduction of the melt pool images, and both the encoder and decoder contain 3 convolutional layers. The AE was pretrained on the image dataset to learn a compressed 512-dimensional latent vector representation of each image. The LSTM uses 4 hidden layers followed by a fully connected regressor head to produce a predicted latent vector. The LSTM processes a sequence of quantum encoded features. The FCNN uses 4 hidden layers to predict the latent vector from a single, quantum encoded feature vector. Hyperparameters for the LSTM and FCNN were tuned to minimize the mean squared error on the validation set.

*4.3.3 K-Means Clustering of Process Parameters*

To manage the significant computation cost of quantum encoding, we use a clustering strategy for the LSTM model. This approach is specific to the LSTM because its sequential architecture allows for the clustering of individual time-step parameter vectors, with many similar parameter vectors. Using mini-batch K-means [25], the process parameter vectors from all time steps are grouped into 50,000 representative cluster centers.

Given the more than 40 billion total time steps in the dataset, K-Means reduces the number of required quantum circuit executions by a factor of approximately 800,000. During network training and inference, the precomputed quantum encoding of a data point's assigned cluster center is used instead of the exact process parameter encoding.

## 4.4 Simulation Results

We first evaluate the performance using a noiseless simulator (Qiskit Estimator). The hybrid quantum-classical models are benchmarked against their purely classical counterparts, which received the raw process parameters as input. Performance is assessed by comparing the predicted melt pool image against the ground truth using mean squared error (MSE), peak signal-to-noise-ratio (PSNR), and structural similarity index metric (SSIM).

**Table 2:** Melt pool prediction results using only classical inputs and using noiseless quantum encoded inputs.

| | Network Type | Evaluation Metric<br>MSE (↓) | PSNR (↑) | SSIM (↑) |
|---|---|---|---|---|
| LSTM | Classical | $1.641 \times 10^{-3}$ | 35.23 | 0.9208 |
| | Quantum | $1.588 \times 10^{-3}$ | 35.00 | 0.9220 |
| | Quantum (K-means) | $1.608 \times 10^{-3}$ | 35.00 | 0.9213 |
| FCNN | Classical | $1.744 \times 10^{-3}$ | 34.01 | 0.8154 |
| | Quantum | $1.681 \times 10^{-3}$ | 34.55 | 0.9185 |

The results, shown in Table 2, indicate the viability and potential benefit of quantum feature encodings. First, both the LSTM and FCNN networks show almost universal improvement when augmented with the quantum feature encoder. While modest, these improvements serve as a successful proof-of-concept for the hybrid quantum-classical methodology. Since the goals of this paper are to explore the potential use of the NISQ processors for LPBF applications, we leave optimization of the quantum circuit to improve performance as future work. Second, the results show the LSTM architecture outperforms the FCNN, highlighting the value of modeling the temporal sequences of the process parameters. Third, the LSTM model trained using features from the K-means cluster centers shows only a small decrease in performance, validating the K-means clustering method as an effective and computationally efficient strategy. Given the computational speedup provided by K-means, the LSTM network with K-means encoding is selected for all subsequent analyses.

Beyond noiseless simulations, we investigate the impact of statistical noise that arises from finite measurement shots, a primary source of noise in NISQ devices for observable measurements. Using the Qiskit Sampler primitive, separate LSTM models are trained on K-means cluster centers encoded with varying shot counts (16 to 4096). Each model is then evaluated on test sets generated with every shot count, with the results shown in Figure 6.

The analysis reveals a surprising insight: prediction accuracy is predominantly determined by the consistency of the shot noise between training and testing, not the absolute number of measurement shots. As shown in Figure 6, models evaluated with the same number of shots used during their training show a consistently low MSE, regardless of the shot count. Even though the sampled data may be noisy for small counts, the quantum encoder provides enough separation between different process parameters to distinguish different features.

The performance of the network degrades when training and evaluation shot counts are mismatched. The performance drop is asymmetric: the model responds differently to an increase versus a decrease in the number of evaluation shots relative to the training shots. When a network is tested with more shots than trained on, the MSE increases but is constant for all larger shots. Meanwhile, when tested with fewer shots, the MSE increases rapidly as fewer shots are used. A possible explanation is that using fewer shots adds noise, while using more shots reduces noise. A practical implication of these results is that achieving high performance does not require many measurement shots, provided that the shot count used during inference is consistent with the shot count used for training.

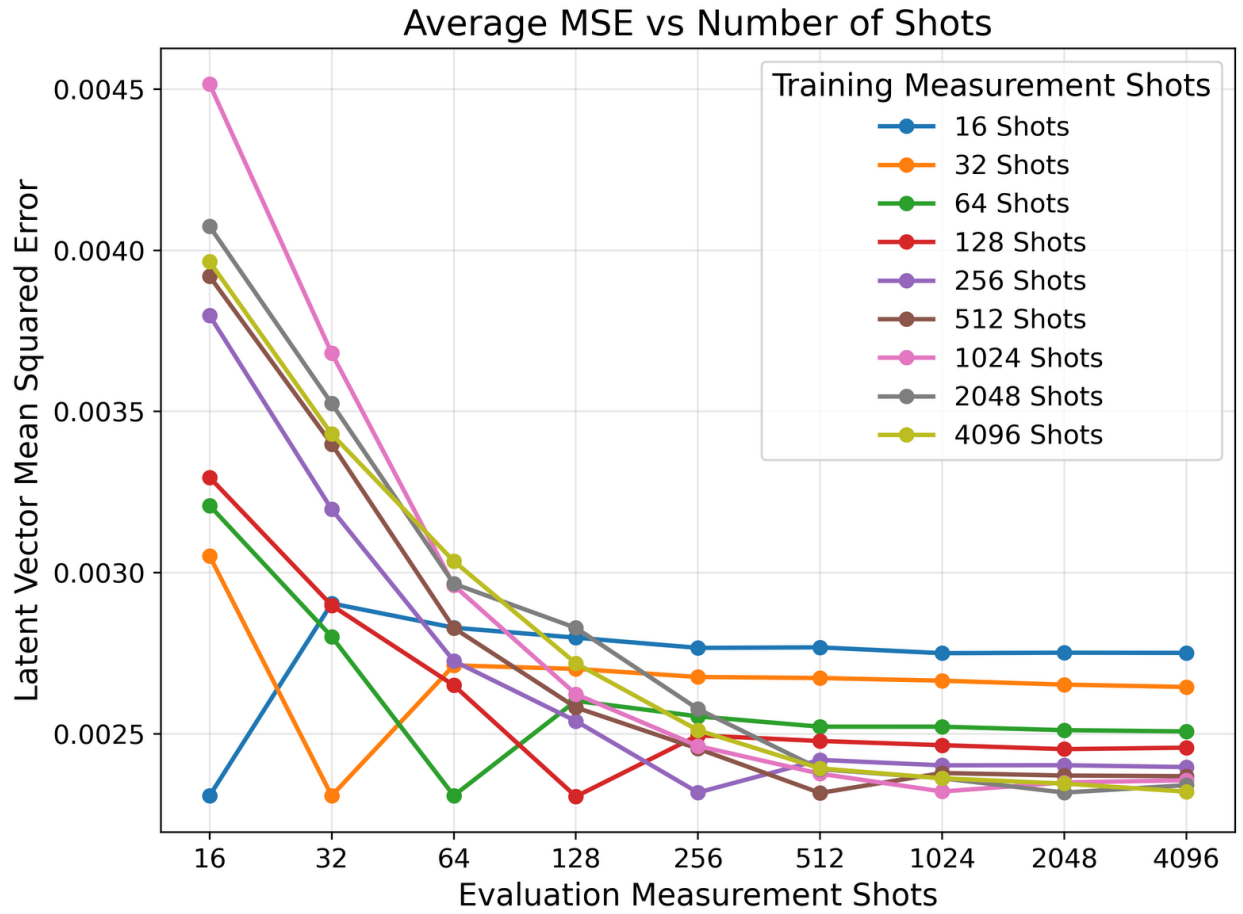

**Figure 6:** The mean squared error for models trained with different measurement shots, as evaluated on data using a variety of measurement shots.

### 4.5 Verification with Quantum Hardware

To complete the validation of our approach, we transition from simulation to physical quantum hardware, executing the circuit on the 133-qubit IBM Torino processor. Evaluating on hardware tests the method's robustness to real-world noise, consisting of both statistical shot noise but also quantum control noise from gate errors and decoherence. Due to practical constraints on the cost of quantum compute time, the experiment was performed on a subset of the data. We select the 725 most frequent K-means cluster centers, covering 23% of the total dataset (190,022 training samples). An LSTM network was trained using the features generated by the quantum hardware. For direct comparison, this model was benchmarked against both a fully classical model and the simulated quantum model, each trained on the same 23% data subset.

The results, shown in Table 3, show that the quantum processor achieves higher performance than both the classical approach and the quantum simulator. Despite the additional noise from imperfect quantum control, current NISQ devices generate improved feature vectors for LPBF process parameters. The improved results suggest the hardware noise serves as a form of regularization, encouraging the network to learn the critical features and improving generalization on the test set.

Table 3 includes the quantum encoding times on the IBM Torino and classical network prediction times using a NVIDIA

**Table 3:** Performance comparison between the simulated quantum encoder and IBM Torino quantum processor on the 23% subset of data.

| QPU Type | Computing Time: Encode (s) | Computing Time: Predict (ms) | Evaluation Metric: MSE (↓) $\times 10^{-3}$ | Evaluation Metric: PSNR (↑) | Evaluation Metric: SSIM (↑) |
|---|---|---|---|---|---|
| Classical | 0 | 3.49 | 1.599 | 31.44 | 0.9209 |
| Simulator | ~0.01 | 3.49 | 1.594 | 31.43 | 0.9210 |
| Real | ~3.4 | 3.49 | 1.590 | 31.45 | 0.9212 |

GeForce GTX 1080 Ti GPU (11 GB memory). While the quantum encoding time is significant (~3.4 s), these features can be computed offline with a one-time cost due to our clustering strategy. Quantum precomputation is essential for real-time applications, as only the faster classical network is required for inference at runtime. Further, quantum computers are subject to availability constraints, and precomputation mitigates the challenge of limited hardware availability. By decoupling quantum encoding from rapid classical prediction, our hybrid approach remains practical for real-time applications.

### 4.6 Discussion

To gain insight into the representational capabilities of the quantum encoder, we analyze the generated feature space using permutation feature importance [32]. The analysis is done on the simulated K-Means LSTM model to identify which quantum observables are most influential in prediction of the melt pool latent vector. The results, shown in Figure 7, demonstrate a distinct pattern. The most impactful observables ($\sigma_7^y$ and $\sigma_4^x$) are characterized by broadly distributed expectation values, indicating these observables are sensitive to variation in the input process parameters. Meanwhile, observables with low importance (illustrated by $\sigma_{10}^z$ and $\sigma_0^z$) exhibit histograms with a sharp peak, suggesting they provide less discriminative information to the neural network. Thus, the quantum encoder is most effective when it creates a varied and responsive feature mapping. The distributed features also explain why the quantum encoder improves the prediction results. By first mapping the classical data into the exponentially large quantum Hilbert space, the data becomes naturally separated which allows the classical neural network to improve melt pool predictions.

Importantly, the analysis of feature importance indicates a direct link between the quantum circuit design and the quality of the quantum encoded features. We observe that observables corresponding to the first and last qubits in the linear chain rank low in importance. We hypothesize that the low importance is a direct consequence of the circuit's linear connectivity, which results in weaker entanglement for outer qubits than for the inner qubits. Therefore, future work can investigate alternative circuit topologies that create stronger and more uniform entanglement to improve quantum feature encoders.

## 5. Conclusion

This paper presents and validates a hybrid quantum-classical neural network approach for predicting LPBF melt pool morphology. Our approach uses a parameterized quantum circuit to encode LPBF process parameters into the high-dimension Hilbert space, before being measured using quantum observables. The primary contributions of the work are fourfold. First, we demonstrate that quantum encoded features can lead to improvements in prediction accuracy for both LSTM and FCNN networks compared to purely classical approaches. Second, we propose K-means clustering to drastically reduce the number of quantum circuit executions without a significant loss in performance. Third, our analysis of measurement shot noise reveals that prediction accuracy is governed by the consistency of noise between training and inference, not the number of shots. Finally, we verify these findings on quantum hardware, demonstrating that the performance of current NISQ processors is sufficient for LPBF applications, achieving results on par with noiseless simulation.

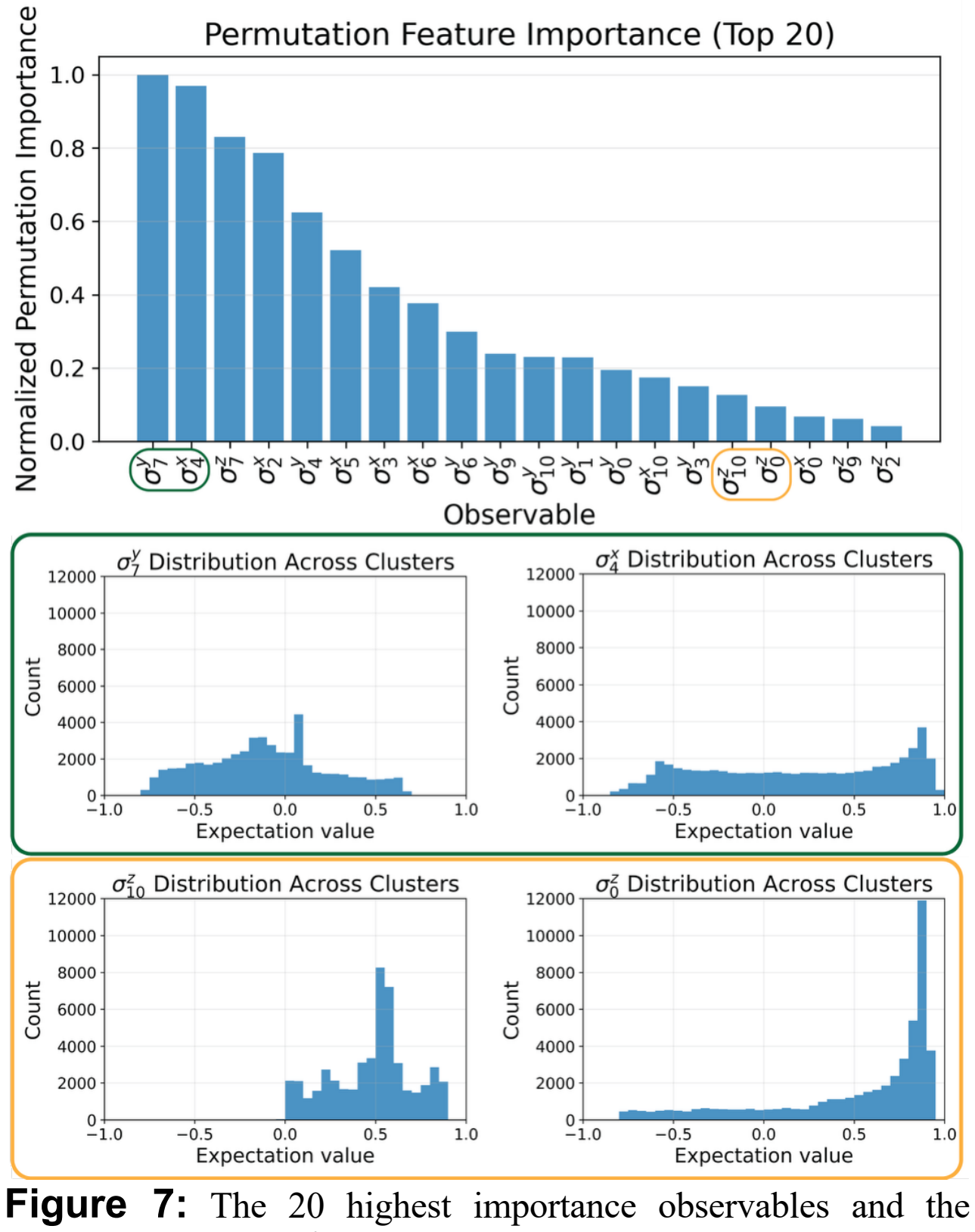

**Figure 7:** The 20 highest importance observables and the histograms for the $\sigma_7^y$, $\sigma_4^x$, $\sigma_{10}^z$, and $\sigma_0^z$ observables.

This work establishes a foundation for applying quantum machine learning to data-driven process monitoring and control in LPBF. Future work should prioritize the optimization of the quantum encoder, aiming to improve the marginal performance improvements demonstrated in this paper. Specifically, future work may explore alternative quantum circuit designs and Hamiltonians. In particular, the relationship between entanglement and feature quality can be investigated, as our initial analysis suggests that circuit topology plays a critical role in neural network performance. Further, the effect of input parameter ordering can be studied: while a classical fully connected layer is permutation-invariant, the structure of our quantum circuit is not. Beyond feature encoding, an alternative approach may explore end-to-end neural network training with a variational quantum algorithm.

In conclusion, this research provides a tangible demonstration of quantum-enhanced machine learning in a practical engineering domain. By successfully employing a NISQ processor to extract more expressive features from LPBF process data, we show a path forward for leveraging the unique representational power of quantum computers. This hybrid approach represents a promising step towards building more accurate data-driven models for quality assurance in AM and highlights a near-term application where quantum computing can provide value.

## Acknowledgement

This material is based upon work supported by the National Science Foundation (NSF) Graduate Research Fellowship Program under Grant No. DGE-2146755. Any opinions, findings, and conclusions or recommendations expressed in this material are those of the authors and do not necessarily reflect the views of the NSF.

## References

[1] Chia, H. Y., Wu, J., Wang, X., and Yan, W. "Process parameter optimization of metal additive manufacturing: a review and outlook." *Journal of Materials Informatics* Vol. 2 No. 4 (2022): pp. 16. DOI 10.20517/jmi.2022.18.

[2] du Plessis, A. "Effects of process parameters on porosity in laser powder bed fusion revealed by X-ray tomography." *Additive Manufacturing* Vol. 30 (2019): pp. 100871. DOI 10.1016/j.addma.2019.100871.

[3] McCann, R., Obeidi, M. A., Hughes, C., McCarthy, É., Egan, D. S., et al. "In-situ sensing, process monitoring and machine control in Laser Powder Bed Fusion: A review." *Additive Manufacturing* Vol. 45 (2021): pp. 102058. DOI 10.1016/j.addma.2021.102058.

[4] Taherkhani, K., Ero, O., Liravi, F., Toorandaz, S., and Toyserkani, E. "On the application of in-situ monitoring systems and machine learning algorithms for developing quality assurance platforms in laser powder bed fusion: A review." *Journal of Manufacturing Processes* Vol. 99 (2023): pp. 848–897. DOI 10.1016/j.jmapro.2023.05.048.

[5] Acharya, R., Abanin, D. A., Aghababaie-Beni, L., Aleiner, I., Andersen, T. I., et al. "Quantum error correction below the surface code threshold." *Nature* Vol. 638 No. 8052 (2025): pp. 920–926. DOI 10.1038/s41586-024-08449-y.

[6] Kim, Y., Eddins, A., Anand, S., Wei, K. X., van den Berg, E., et al. "Evidence for the utility of quantum computing before fault tolerance." *Nature* Vol. 618 No. 7965 (2023): pp. 500–505. DOI 10.1038/s41586-023-06096-3.

[7] Nielsen, M. A. and Chuang, I. L. *Quantum Computation and Quantum Information: 10th Anniversary Edition*. Cambridge University Press (2010).

[8] Liu, J., Ye, J., Silva Izquierdo, D., Vinel, A., Shamsaei, N., et al. "A review of machine learning techniques for process and performance optimization in laser beam powder bed

fusion additive manufacturing." *Journal of Intelligent Manufacturing* Vol. 34 No. 8 (2023): pp. 3249–3275. DOI 10.1007/s10845-022-02012-0.

[9] Preskill, J. "Quantum Computing in the NISQ era and beyond." *Quantum* Vol. 2 (2018): pp. 79. DOI 10.22331/q-2018-08-06-79.

[10] Reddy, K. P. K., Rao, B. N., Nazeemudheen, M. N., Manwatkar, S. K., and Murty, S. V. S. N. "Selection of Optimal Process Parameters to Obtain Defect-Free Builds in IN718 Made by Laser Powder Bed Fusion." *Journal of Materials Engineering and Performance* Vol. 33 No. 19 (2024): pp. 10228–10241. DOI 10.1007/s11665-023-08677-9.

[11] Xu, S., Younes Araghi, M., and Su, Y. "Physics-based modeling of metal additive manufacturing processes: a review." *The International Journal of Advanced Manufacturing Technology* Vol. 134 No. 1 (2024): pp. 1–13. DOI 10.1007/s00170-024-14156-9.

[12] Sato, M. M., Wong, V. W. H., Yeung, H., Witherell, P., and Law, K. H. "Identification and Interpretation of Melt Pool Shapes in Laser Powder Bed Fusion with Machine Learning." *Smart and Sustainable Manufacturing Systems* Vol. 8 No. 1 (2024): pp. 1–23. DOI 10.1520/SSMS20230035.

[13] Mohammed, A. S., Almutahhar, M., Sattar, K., Alhajeri, A., Nazir, A., et al. "Deep learning based porosity prediction for additively manufactured laser powder-bed fusion parts." *Journal of Materials Research and Technology* Vol. 27 (2023): pp. 7330–7335. DOI 10.1016/j.jmrt.2023.11.130.

[14] Cao, Y., Chen, C., Xu, S., Zhao, R., Guo, K., et al. "Machine learning assisted prediction and optimization of mechanical properties for laser powder bed fusion of Ti6Al4V alloy." *Additive Manufacturing* Vol. 91 (2024): pp. 104341. DOI 10.1016/j.addma.2024.104341.

[15] Biamonte, J., Wittek, P., Pancotti, N., Rebentrost, P., Wiebe, N., et al. "Quantum machine learning." *Nature* Vol. 549 No. 7671 (2017): pp. 195–202. DOI 10.1038/nature23474.

[16] Schuld, M. and Killoran, N. "Quantum Machine Learning in Feature Hilbert Spaces." *Physical Review Letters* Vol. 122 No. 4 (2019): pp. 040504. DOI 10.1103/PhysRevLett.122.040504.

[17] Cerezo, M., Arrasmith, A., Babbush, R., Benjamin, S. C., Endo, S., et al. "Variational quantum algorithms." *Nature Reviews Physics* Vol. 3 No. 9 (2021): pp. 625–644. DOI 10.1038/s42254-021-00348-9.

[18] Jaksch, D., Givi, P., Daley, A. J., and Rung, T. "Variational Quantum Algorithms for Computational Fluid Dynamics." *AIAA Journal* Vol. 61 No. 5 (2023): pp. 1885–1894. DOI 10.2514/1.J062426.

[19] Chen, S. Y.-C., Huang, C.-M., Hsing, C.-W., Goan, H.-S., and Kao, Y.-J. "Variational quantum reinforcement learning via evolutionary optimization." *Machine Learning: Science and Technology* Vol. 3 No. 1 (2022): pp. 015025. DOI 10.1088/2632-2153/ac4559.

[20] Larocca, M., Thanasilp, S., Wang, S., Sharma, K., Biamonte, J., et al. "Barren plateaus in variational quantum computing." *Nature Reviews Physics* Vol. 7 No. 4 (2025): pp. 174–189. DOI 10.1038/s42254-025-00813-9.

[21] Ciceri, A., Cottrell, A., Freeland, J., Fry, D., Hirai, H., et al. "Enhanced fill probability estimates in institutional algorithmic bond trading using statistical learning algorithms with quantum computers." (2025): DOI 10.48550/arXiv.2509.17715.

[22] Choi, E., Sul, J., Kim, J. E., Hong, S., Gonzalez, B. I., et al. "Quantum machine learning for additive manufacturing process monitoring." *Manufacturing Letters* Vol. 41 (2024): pp. 1415–1422. DOI 10.1016/j.mfglet.2024.09.168.

[23] Heisenberg, W. "Zur Theorie des Ferromagnetismus." *Zeitschrift für Physik* Vol. 49 No. 9 (1928): pp. 619–636. DOI 10.1007/BF01328601.

[24] Suzuki, M. "Generalized Trotter's formula and systematic approximants of exponential operators and inner derivations with applications to many-body problems." *Communications in Mathematical Physics* Vol. 51 No. 2 (1976): pp. 183–190. DOI 10.1007/BF01609348.

[25] Sculley, D. "Web-Scale K-Means Clustering." *Proceedings of the 19th international conference on World wide web* (2010): pp. 1177–1178. DOI 10.1145/1772690.1772862.

[26] Lane, B. M. and Yeung, H. "Process Monitoring Dataset from the Additive Manufacturing Metrology Testbed (AMMT): 'Three-Dimensional Scan Strategies.'" *NIST Journal of Research* (2019): DOI 10.6028/jres.124.033.

[27] Lane, B. and Yeung, H. "Process Monitoring Dataset from the Additive Manufacturing Metrology Testbed (AMMT): 3D Scan Strategies." https://data.nist.gov/od/id/85196AB9232E7202E053245706813DFA2044.

[28] Javadi-Abhari, A., Treinish, M., Krsulich, K., Wood, C. J., Lishman, J., et al. "Quantum computing with Qiskit." *arXiv* (2024): DOI 10.48550/arXiv.2405.08810.

[29] Vidal, G. "Efficient Classical Simulation of Slightly Entangled Quantum Computations." *Physical Review Letters* Vol. 91 No. 14 (2003): pp. 147902. DOI 10.1103/PhysRevLett.91.147902.

[30] IBM Quantum, Processor types. Accessed February 13, 2026. https://quantum.cloud.ibm.com/docs/en/guides/processor-types.

[31] Wallman, J. J. and Emerson, J. "Noise tailoring for scalable quantum computation via randomized compiling." *Physical Review A* Vol. 94 No. 5 (2016): pp. 052325. DOI 10.1103/PhysRevA.94.052325.

[32] Breiman, L. "Random Forests." *Machine Learning* Vol. 45 No. 1 (2001): pp. 5–32. DOI 10.1023/A:1010933404324.